\documentclass[ twocolumn]{aastex62}
\hypersetup{linkcolor=red,citecolor=blue,filecolor=cyan,urlcolor=magenta}
\usepackage{filecontents}
\usepackage{booktabs}
\usepackage{amsmath}
\usepackage{amsfonts}
\usepackage{amssymb}
\usepackage{xcolor}

\newcommand{\si}{$\xi_{ion}$}

\newcommand{\one}{\textit{Standard}}
\newcommand{\two}{\textit{Effective}}
\newcommand{\oiii}{[$\mathrm{O_{III}}$]}
\newcommand{\uv}{$\mathrm{L_{UV}}$}

\begin{document}
\font\myfont=cmr12 at 14pt

\title{ THE IONIZING PHOTON PRODUCTION EFFICIENCY ($\xi_{ion}$) OF LENSED DWARF GALAXIES AT $z\sim 2 $ \footnote{Some of the data presented herein were obtained at the W. M. Keck Observatory, which is operated as a scientific partnership among the California Institute of Technology, the University of California and the National Aeronautics and Space Administration. The Observatory was made possible by the generous financial support of the W. M. Keck Foundation. \\
This research is based on observations made with the NASA/ESA Hubble Space Telescope obtained from the Space Telescope Science Institute, which is operated by the Association of Universities for Research in Astronomy, Inc., under NASA contract NAS 5?26555. These observations are associated with programs 12201, 12931, 13389 and 14209.
}}
\correspondingauthor{Najmeh Emami}
\email{najmeh.emami@email.ucr.edu}
\author{ Najmeh Emami}
\affiliation{ Department of Physics and Astronomy, University of California Riverside, Riverside, CA 92521, USA}

\author{ Brian Siana}
\affiliation{ Department of Physics and Astronomy, University of California Riverside, Riverside, CA 92521, USA}

\author{ Anahita Alavi}
\affiliation{ Infrared Processing and Analysis Center, Caltech, Pasadena, CA 91125, USA}

\author{Timothy Gburek}
\affiliation{ Department of Physics and Astronomy, University of California Riverside, Riverside, CA 92521, USA}

\author{ William R. Freeman}
\affiliation{ Department of Physics and Astronomy, University of California Riverside, Riverside, CA 92521, USA}

\author{Johan Richard}
\affiliation{Univ Lyon, Univ Lyon1, Ens de Lyon, CNRS, Centre de Recherche Astrophysique de Lyon UMR5574, F-69230, Saint-Genis-Laval, France
}

\author{Daniel R. Weisz}
\affiliation{Department of Astronomy, University of California Berkeley, Berkeley, CA 94720, USA}

\author{ Daniel P. Stark}
\affiliation{ Steward Observatory, University of Arizona, 933 N Cherry Ave, Tucson, AZ 85721 USA
}

\begin{abstract}

We measure the ionizing photon production efficiency ($\xi_{ion}$) of low-mass galaxies ($10^{7.8}$-$10^{9.8}$ $M_{\odot}$) at $1.4<z<2.7$ to better understand the contribution of dwarf galaxies to the ionizing background and reionization. We target galaxies that are magnified by strong lensing galaxy clusters and use Keck/MOSFIRE to measure nebular emission line fluxes and {\it HST} to measure the rest-UV and rest-optical photometry. We present two methods of stacking. First, we take the average of the log of H$\alpha$-to-UV luminosity ratios (L$_{H\alpha}$/L$_{UV}$) of galaxies to determine the standard log($\xi_{ion}$). Second, we take the logarithm of the total L$_{H\alpha}$ over the total L$_{UV}$. We prefer the latter as it provides the total ionizing UV luminosity density of galaxies when multiplied by the non-ionizing UV luminosity density. log($\xi_{ion}$) calculated from the second method is $\sim$ 0.2 dex higher than the first method. We do not find any strong dependence between log($\xi_{ion}$) and stellar mass, far-UV magnitude (M$_{UV}$), or UV spectral slope ($\beta$). We report a value of log($\xi_{ion}$) $\sim25.47\pm 0.09$ for our UV-complete sample ($-22<M_{UV}<-17.3$) and $\sim25.37\pm0.11$ for our mass-complete sample ($7.8<\text{log}(M_*)<9.8)$. These values are consistent with measurements of more massive, more luminous galaxies in other high-redshift studies that use the same stacking technique. Our log($\xi_{ion}$) is $0.2-0.3$ dex higher than low-redshift galaxies of similar mass, indicating an evolution in the stellar properties, possibly due to metallicity or age. We also find a correlation between log($\xi_{ion}$) and the equivalent widths of H$\alpha$ and [OIII]$\lambda$5007 fluxes, confirming that these equivalent widths can be used to estimate $\xi_{ion}$.

\end{abstract}

\keywords{galaxies: dwarf --- galaxies: evolution --- galaxies: high-redshift}

\section{Introduction }
\label{sec:intro}

Many studies have demonstrated that by $z\sim6$ the neutral hydrogen in the intergalactic medium (IGM) was mostly ionized \citep{Becker_2001, Fan_2006, McGreer_2015}. What is not well understood is what are the sources that ionized the universe and provided the intergalactic medium thereafter \citep{Fan_2001,  Somerville_2003, Madau_2004, Bouwens_2015A}. In fact, it is not clear whether the galaxies that we have detected at high redshift are capable of ionizing the IGM \citep{Robertson_2015, Finkelstein_2019, Naidu_2019, Mason_2019}. In order to determine this, we need to know the rate of ionizing photons emitted into the IGM as a function of redshift (often referred to as $\Gamma(z)$). In order to calculate $\Gamma(z)$, three quantities must be known.

\begin{equation}
    \Gamma = \int L\Phi(L)\xi_{ion}(L)f_{esc}(L)dL
\end{equation}

The first quantity is the luminosity function of galaxies, $\Phi(L)$, which is typically measured in the non-ionizing ultraviolet (UV), as it is relatively easy to detect galaxies at those wavelengths at high redshift. If the UV luminosity function is integrated, it gives the total non-ionizing UV luminosity density at a given redshift. The second quantity that is needed is a conversion from the non-ionizing UV luminosity density to ionizing UV luminosity density. This conversion is often referred to as \si\ and is defined as the rate of ionizing photon production normalized by the non-ionizing UV luminosity density (in $f_{\nu}$). The third necessary quantity is the fraction of ionizing photons that escape into the intergalactic medium, referred to as the escape fraction, $f_{esc}$. Of course, all of these quantities can vary with luminosity, stellar mass, age, and metallicity.

Many studies have constrained the luminosity functions \citep{ Bouwens_2006, Bouwens_2007, Reddy_2009, Oesch_2013, Alavi_2014, Bouwens_2015B, Mehta_2017} and escape fractions \citep{2006_Inoue, Siana_2007, Wise_2009, Vanzella_2010, Vasei_2016, Japelj_2017, Grazian_2017} of high redshift galaxies . Here we are interested in constraining the second quantity, \si. 
The primary way of determining \si\ is to infer the ionizing UV flux from the hydrogen recombination lines (e.g., $H\alpha$ or $H\beta$) assuming that the interstellar medium (ISM) is optically thick to ionizing photons and does not allow them to escape the galaxy. In this case, the rate of ionizations and, thus, the ionizing photon production rate, can be inferred from recombination lines assuming case-B recombination. As such, \citet{ Bouwens_2016a, Nakajima_2016, Matthee_2017, Shivaei_2018, Tang_2019} evaluated \si\ as the ratio of hydrogen recombination lines to 1500 $\AA$ UV fluxes. Another indirect way of inferring \si\ is to implement metal nebular emission lines and stellar continuum into the photoionization models and output the shape of the ionizing spectrum and, thus, the best \si\ match to the observed spectrum \citep{Stark_2015, Stark_2017, Chevallard_2018}.

However, all of these studies measure \si\ of high-redshift galaxies that are exclusively luminous H$\alpha$ or Ly$\alpha$ emitters or have extreme optical nebular emission lines. As such there are not many measurements of \si\ in low-luminosity, low-mass galaxies \citep{Lam_2019}.

It is not clear what type of galaxies contribute the most to the total ionizing photon budget necessary for reionization. Some studies suggest that perhaps rare Lyman continuum leakers with substantial star-formation surface densities have led to a rapid, recent reionization at z$\sim$6 \citep{Naidu_2019}. Other studies predict that low-mass galaxies should have a greater contribution to reionization because of the steep faint end slope of the UV$_{1500}$ luminosity function of high redshift galaxies \citep{Reddy_2009, Bouwens_2012, Alavi_2014, Finkelstein_2015, Ishigaki_2015, Atek_2015, Livermore_2017, Mehta_2017}. Additionally, at low mass, more ionizing photons are thought to escape from the galaxies into the IGM \citep{Paardekooper_2013, Wise_2014, Erb_2015, Anderson_2017, Henry_2015,Karman_2017} at high redshifts, possibly through hot ``chimneys'' created by feedback-driven outflows. In order to determine whether low-mass galaxies are the primary reionizing agents, we still need to investigate the ionizing photon production efficiency (\si) of these low mass galaxies and compare to their massive counterparts. However, despite its great importance, little is known about the \si\ in faint low mass systems.

In this paper we measure, for the first time, \si\ for low-mass ($7.8 \leq log (M_*) < 9.8$), low-luminosity ($-22<M_{UV}<-17.3$) galaxies at $1.4<z<2.7$. These galaxies may be intermediate-redshift analogs of the sources of reionization at $z>6$. Galaxies in our sample are highly magnified by gravitational lensing by foreground galaxy clusters. The magnification enables us to detect low luminosity galaxies, up to an intrinsic UV magnitude of -17. We quantify \si\ using $H\alpha$ recombination emission and non-ionizing (1500 $\AA$) UV fluxes from deep Keck/MOSFIRE spectroscopy and {\it HST} imaging, respectively. We also have $H\beta$ detections for all of the galaxies in our sample which allows us to correct $H\alpha$ fluxes for the dust extinction via the Balmer decrement. We correct the UV stellar continuum using the dust extinction inferred from the SED fitting. We carefully select galaxies to be complete in both low and high UV luminosities.

There is an intrinsic scatter in the ratio of $H\alpha$ (or $H\beta$) to $UV$, especially in low-mass galaxies \citep{Lee_2009, Weisz_2012, Dominguez_2015, Guo_2016, Emami_2019}. Many factors are known to contribute to this scatter including bursty star formation, galaxy-to-galaxy dust extinction variation, escape of ionizing photons, varying initial stellar mass function (IMF), different stellar metallicities, and stellar models \citep{Iglesias_2004, Boselli_2009, Lee_2009, Meurer_2009, Weisz_2012, Guo_2016, Emami_2019}. 
As a result, we expect to see a similar scatter in the \si\ distribution, which makes it crucial to come up with an appropriate way of combining the galaxies' fluxes and derive a single \si\ value that properly represents the entire sample.
Here we also address this issue and introduce a new way of stacking $H\alpha$ and UV fluxes that deals with the \si\ scatter in low mass galaxies.

Since \si\ is related to the ionizing radiation intensity of the galaxies, it can also be inferred from other physical quantities that are also dependent on the ionizing radiation intensity, such as UV spectral slope \citep{Robertson_2013, Bouwens_2015A, Duncan_Conselice_2015} and equivalent widths of nebular UV and optical emission lines \citep{Stark_2015, Stark_2017, Chevallard_2018, Tang_2019}. We investigate the relationship between \si\ and these physical quantities in our sample and see if the relations shown by previous works further extend to lower luminosity.

The outline of the paper is as follows. We describe the sample selection and data acquisition in $\mathsection$ \ref{sec:data}. In $\mathsection$ \ref{sec:measurement} we present flux measurements. In $\mathsection$ \ref{sec:stacking} we describe two approaches of stacking fluxes and discuss the relevance of each for the \si\ determination. In $\mathsection$ \ref{sec:results} we show our results and compare them with previous works. We provide physical interpretations explaining our results in conjunction with previous studies in $\mathsection$ \ref{sec:discussion}. Lastly, we conclude with a brief summary in $\mathsection$ \ref{sec:summary}. We assume a $\Lambda$-dominated flat Universe with $\Omega_{\Lambda}$ = 0.7, $\Omega_M$ = 0.3 and H$_0$ = 70 km s$^{-1}$ Mpc$^{-1}$. All magnitudes in this paper are in the AB system \citep{Oke_Gunn_1983} and all equivalent widths are quoted in the rest-frame.

\section{data}
\label{sec:data}

\subsection{HST Data}
Our sample is drawn from a {\it Hubble Space Telescope} ({\it HST}) survey \citep{Alavi_2016} that identifies faint star-forming galaxies at $1<z<3$ behind three lensing clusters -- Abell 1689 and two Hubble Frontier Fields (HFF) clusters, MACS J0717 and MACS J1149 \citep{Lotz_2017}. The data reduction and photometric measurements are discussed in detail in \citet{Alavi_2016}. For galaxies behind Abell 1689, we measure flux in eight photometric bands spanning the observed near-UV and optical. For galaxies behind MACS J0717 and MACS J1149 we measure flux in nine photometric bands spanning the observed near-UV, optical, and near-IR. The near-UV data (program IDs 12201, 12931, 13389) allows for efficient identification of the Lyman break, enabling accurate photometric redshifts at $1<z<3$.

We require a lens model for each cluster to correct for the lensing magnification and derive the intrinsic galaxy properties. As discussed in \citet{Alavi_2016}, for Abell 1689 we use the lens model 
of \citet{Limousin_2007} and for the HFF clusters we use the released models from the  CATS\footnote{https://archive.stsci.edu/prepds/frontier/lensmodels/} team (\citet{2016_Jauzac} and \citet{Limousin_2016} for MACS J1149 and MACS J0717, respectively). According to \citet{Priewe_2017}, for the HFF clusters and a typical magnification of our galaxies which is around 5, the systematic error in the estimated magnification of different lens models is  $\sim40$\%, which is small compared to our $M_*$ or UV luminosity bin sizes which are about 1 order of magnitude (Figures \ref{fig:xi_mass_Weisz_Shivaei} and \ref{fig:xi_Muv}). In \citet{Alavi_2016} there is a full description of the lens models used for the A1689 and HFF clusters.

\subsection{Spectroscopic Sample and Data Reduction}
We obtain the rest-frame optical spectra of our sample via Keck/MOSFIRE observation.
We select our spectroscopic sample such that the 
bright rest-frame optical nebular emission lines fall within the atmospheric windows at $1.37<z<1.70$ and $2.09<z<2.61$. When selecting targets, we prioritized galaxies with high magnification and brighter observed optical flux densities ($M_B<26.5$). The data were collected between January 2014 and March 2017. Masks were made for the $1.37<z<1.70$ and $2.09<z<2.61$ redshift ranges and all of the strong optical emission lines (H$\alpha$, [N{\sc ii}], [O{\sc iii}], H$\beta$, and [O{\sc ii}]) were targeted. For the lower redshift mask, Y-, J-, and H-band spectroscopy was obtained. For the higher redshift mask, J-, H-, and K-band spectroscopy was obtained. The total exposure times for each mask and filter range from 48 to 120 minutes. The typical FWHM seeing of our MOSFIRE spectra in any given mask and filter is $\sim 0.71''$. The slit widths are also $0.70''$.

The MOSFIRE data were reduced using the MOSFIRE Data Reduction Pipeline\footnote{https://keck-datareductionpipelines.github.io/MosfireDRP/}(DRP). The DRP produces a 2D flat-fielded, sky-subtracted, wavelength-calibrated, and rectified spectrum for each slit. It also combines the spectra taken at each nod position (we used an ABBA dither pattern). The wavelength calibration for the J- and H-band spectra was performed using the skylines and for the Y- and K-band spectra a combination of skylines and Neon lines. 
We then utilize custom IDL software, BMEP \footnote{https://github.com/billfreeman44/bmep}, from \citet{Freeman_2019} for the 1D extraction of spectra.
The flux calibration is done in two stages. First, we use a standard star with spectral type ranging from B9 V to A2 V, which has been observed at similar airmass as the mask, to determine a wavelength-dependent flux calibration. We then do an absolute flux calibration using the spectrum of a star to which we assigned a slit in each mask.

\section{Measurements}
\label{sec:measurement}

\subsection{SED Fitting}
\label{subsec:sed}
Stellar masses, star formation rates and stellar dust attenuation for our galaxies are estimated with SED fits to the photometry.
Specifically, for the Abell 1689 cluster, we use eight broad-band filters spanning the observed near-UV to optical in the F225W, F275W, F336W, F475W, F625W, F775W, F814W and F850LP filters. In addition, we use the photometry in two near-IR HST bands (F125W and F160W), though the imaging does not cover the full area covered by the near-UV and optical imaging.

For the two HFF clusters, we fit to nine broad-band filters spanning the observed near-UV to near-IR in the F275W, F336W, F435W, F606W, F814W, F105W, F125W, F1140W and F160W filters. 

We use the stellar population fitting code FAST \citep{Kriek_2009}, with the BC03 \citep{Bruzual_2003} population synthesis models, and assume an exponentially increasing star formation history (which has been shown to best reproduce the observed SFRs at z $\sim$ 2; \citealt{Reddy_2012}) with a Chabrier IMF \citep{Chabrier_2003}. 
As suggested by \citet{Reddy_2018} for high-redshift low-mass galaxies, we use the SMC dust extinction curve \citep{Gordon_2003} with $A_V$ values varying between $0.0-3.0$. We leave the metallicity as a free parameter between [0.4-0.8] Z$_{\odot}$. The age and star formation timescales can vary between $7.0<\log(t) \ [$yr$]<10$ and $8.0<\log(\tau) \ [$yr$]<11.0$, respectively. The redshifts are fixed to the values obtained spectroscopically. The $1\sigma$ confidence intervals are derived from a Monte Carlo method of perturbing the broad-band photometry within the corresponding photometric uncertainties and refitting the SED 300 times. We note that we correct the broadband photometry for the contamination from the nebular emission lines using the line fluxes measured from the MOSFIRE spectra. 

\subsection{Emission Lines Fitting}
\label{line_fitting}
Spectral fitting was performed in each filter covering a galaxy, and for all of the aforementioned strong rest-optical emission lines, using the Markov Chain Monte Carlo (MCMC) Ensemble sampler, \texttt{emcee}\footnote{\url{https://emcee.readthedocs.io/en/v2.2.1/}} \citep{Foreman-Mackey_2013}. Before fitting, to account for sky line contamination within a given spectrum, we removed any data points that have a corresponding error $>3\times$ the median error of the spectrum. Emission lines within close proximity to each other in a filter (e.g.; [\ion{O}{3}]$\lambda\lambda$4959,5007 and H$\beta$; [\ion{N}{2}]$\lambda\lambda$6548,6583 and H$\alpha$) were fit simultaneously with single-Gaussian profiles, and the continuum was fit with a line. For the emission lines relevant to this paper (H$\beta$, [\ion{O}{3}]$\lambda$5007, and H$\alpha$), the free parameters of the fits comprised the slope and y-intercept of the continuum line, a single emission-line width ($\sigma$) and redshift for the filter, and the amplitudes ($A_\lambda$) of the individual lines. When fitting the portion of the spectrum containing H$\beta$ and the [\ion{O}{3}] doublet, the amplitude of [\ion{O}{3}]$\lambda$5007 was set as a free parameter with the amplitude of [\ion{O}{3}]$\lambda$4959 constrained to follow the intrinsic flux ratio of the doublet's lines: [\ion{O}{3}]$\lambda$5007/[\ion{O}{3}]$\lambda$4959 = 2.98 \citep{Storey&Zeippen_2000}. In the instances where [\ion{O}{3}]$\lambda$5007 fell outside our spectroscopic coverage, its flux was determined with this flux ratio and the [\ion{O}{3}]$\lambda$4959 line. The final spectroscopic redshift of a galaxy was determined via the weighted average of the redshifts fit to the different filters. More details about the spectroscopic line measurements of [\ion{O}{2}] and other, fainter optical lines not considered here can be found in \citet{Gburek_2019}. To assess the quality of the fits to the spectra, posterior histograms were output for each free parameter (as well as for the fluxes), and 68$\%$ confidence intervals were fit to the histograms. 


\subsection{Slit Loss Correction}
The emission line fluxes need to be corrected for slit losses. This procedure is more important for extended or stretched (highly-magnified) objects as the slit may not fully cover the object. This needs to be done for each object in each MOSFIRE band and each mask. We adopt the following procedures: 1. We cut a 30$''\times$30$''$ postage stamp centered on the galaxy from the F625W as this filter gives a high $S/N$ image of the rest-frame ultraviolet light, and therefore the approximate morphology of the star-forming regions. 2. We identify the pixels corresponding to the object using the segmentation map output by SourceExtractor \citep{SExtractor_1996}. 3. We mask out all pixels of the nearby objects and background from the postage stamp and replace them with zero flux. 4. The sum of the total flux from pixels belonging to the object gives us the actual flux that SourceExtractor measured. 5. We smooth the postage stamp applying a Gaussian kernel with a FWHM that is given by

\begin{equation}
    FWHM^2_{kernel} = FWHM^2_{seeing}- FWHM^2_{F625W}
\end{equation}

$FWHM^2_{seeing}$ is the FWHM of the Gaussian fit to the profile of the slit star in the corresponding mask and filter. $FWHM^2_{F625W}$ is the FWHM of the F625W PSF (0.1$''$). This artificially degrades the resolution of the {\it HST} image to the same spatial resolution as the MOSFIRE observation. 6. We overlay the slit on the postage stamp of the smoothed image using its position angle, center, length and width and block out regions of the object that falls out of the slit. 7. We sum the flux of the remaining pixels in the slit and denote it as in-slit flux. 8. We then determine a multiplicative factor required to have the in-slit flux match the total flux. This factor is the slit loss correction and is applied to all lines in the corresponding filter and mask.

\subsection{Sample Selections}
\label{subsec:sample_select}
There are 62 galaxies in our sample for which we have sufficient {\it HST} filter coverage spanning the observed near-UV to near-IR that enables a robust SED fit and, thus, reliable estimates of stellar properties (stellar mass, V-band dust attenuation ($A_{V}$), UV spectral slope ($\beta$), etc.). We remove some galaxies from the sample for the following reasons:

\noindent\textbf{Non-covered H$\alpha$ or large H$\alpha$ errors:} We remove 16 galaxies that do not have good H$\alpha$ measurements, either because H$\alpha$ is out of the wavelength coverage, or the errors are significantly larger than the H$\alpha$ error distribution of the sample ($>10^{-17}$ erg s$^{-1}$ cm$^{-2}$, corresponding to a typical error $>$ 1.76 $\times$ the median error, typically due to strong sky line contamination). The median H$\alpha$ flux error of the sample is 6.08$ \times 10^{-18}$ erg s$^{-1}$ cm$^{-2}$. We choose to impose a flux error cutoff rather than a signal-to-noise cutoff on our H$\alpha$ measurement, in order to avoid a bias against intrinsically faint H$\alpha$ emitters.

\noindent\textbf{Galaxies with very high magnification:}  If a galaxy has a high average magnification, it means it is sitting close to the caustic in the source plane. Thus, the gradient of the magnification can be large, resulting in large magnification differences across the galaxy. This could result in an observed  ratio of $L_{H\alpha}$ to $L_{UV}$ that is different from the true ratio. Not only would this increase the scatter, but it can result in a bias, as the galaxies are selected via rest-frame UV continuum luminosity density. Hence, we remove 7 galaxies whose magnifications ($\mu$) are $\mu>30 $ in A1689 and $\mu>15 $ in HFF clusters. The reason for choosing a larger magnification cut for the A1689 compared to other HFF clusters is because A1689 has a large Einstein radius, which provides high magnification over a large area in the source plane. Therefore, objects with a high magnification in A1689 are not required to be close to the critical lines, where the magnification formally diverges \citep{Alavi_2016}.

\noindent\textbf{Multiply-imaged galaxies}  We remove multiple images of two galaxies to avoid double-counting. In these cases, we keep the most highly magnified image in the sample unless the magnification is very large ($>30$ in A1689 and $>15$ in HFF clusters), in which case, we use the next brightest image. These multiple images were identified using Lenstool \citep{Limousin_2016,Alavi_2016}.

\noindent\textbf{High slit-loss galaxies:} For larger, extended galaxies, the slit loss correction can be large, and the MOSFIRE measurement will only be sampling a small, possibly unrepresentative portion of the whole galaxy.

As such, we remove four galaxies with $H\alpha$ slit losses $>70$\% from the sample. This is also worth noting that the typical slit loss of our sample is 40\%.

\noindent\textbf{Galaxies with large mass errors:} We also make sure not to include galaxies with large stellar mass errors in our analysis. There are only four galaxies that lack {\it HST} rest-frame near-IR filter coverage and ultimately end up having large mass errors shown as gray points in Figure \ref{fig:all_sample}. We note that we only exclude these galaxies from our sample when we perform flux stacking based on stellar masses (Section \ref{subsec:xi_mass}), but use them when stacking based on properties other than the stellar mass (UV magnitude and UV spectral slope, Sections \ref{subsec:xi_Muv} and \ref{subsec:xi_beta}).

The final sample contains 28 galaxies that are free of the aforementioned concerns.

\subsection{Non-Dust-Corrected \si}
\label{subsec:non_dust_corr_xi}

The goal of this paper is to measure the ionizing photon production efficiency of galaxies ($\xi_{ion}$) for our sample, which represents the rate of Lyman continuum photons per unit UV$_{1500}$ luminosity as:

\begin{equation}
    \xi_{ion}=\frac{Q_{H^0}}{L_{UV}}\ [s^{-1}/erg.\ s^{-1}.\ Hz^{-1}]
\end{equation}

{\noindent}where $L_{UV}$ is the intrinsic UV-continuum luminosity density (per unit frequency) around $1500$ \AA. Based on Case-B recombination, the rate of production of ionizing photons ($Q_{H^0}$) can be determined from the hydrogen recombination lines, in this case $H\alpha$, as

\begin{equation}
    L_{H\alpha}[erg.\ s^{-1}]=1.36\times 10^{-12}\ Q_{H^0}[s^{-1}]
\end{equation}

{\noindent}where $ L_{H\alpha}$ is the $H\alpha$ luminosity \citep{Leitherer_1995}. Here we assume that all ionizing photons result in a photoionization (none escape into the IGM) and are converted into case B recombination emission. Therefore, the \si\ values reported in this study are upper limits.

\begin{figure}
    \centering
    \includegraphics[width=1 \linewidth]{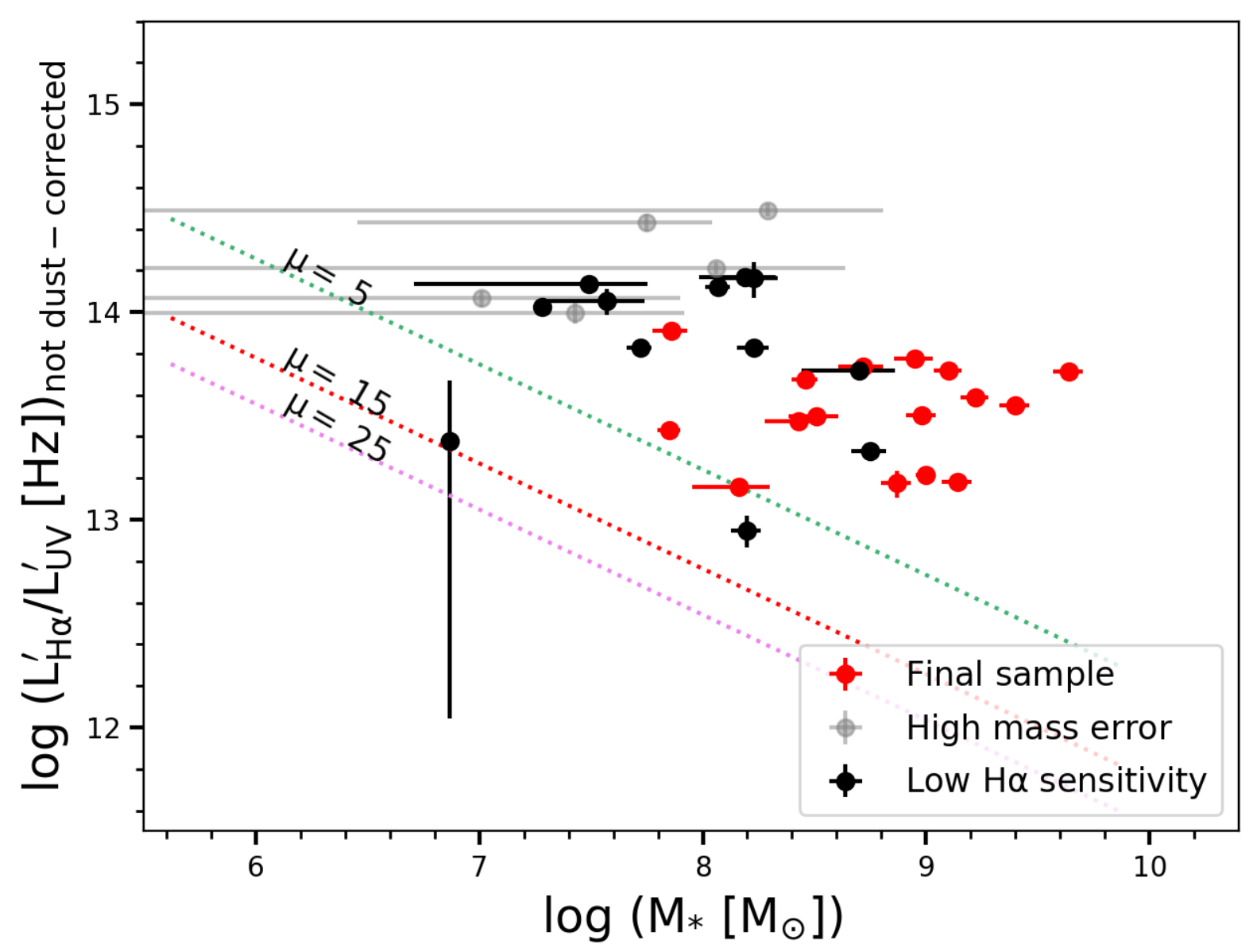}
    \caption{Not dust-corrected log($L_{H\alpha}^{\prime}/L_{UV}^{\prime}$) as a function of log(M$_*$) derived from the observed $L_{H\alpha}$ and $L_{UV}$. The gray points show galaxies with high mass errors. Black points indicate galaxies that could not be detected if they had the very low observed log ($L_{H\alpha}^{\prime}/L_{UV}^{\prime}$)$<13.2$. The green, red and magenta diagonal dotted lines indicate the typical log($L_{H\alpha}^{\prime}/L_{UV}^{\prime}$) detection limit for three magnification factors of 5, 15 and 25 respectively, below which galaxies are intrinsically too faint to be detected through MOSFIRE. The remaining galaxies in red are free of any biases in our measurements.}
    \label{fig:all_sample}
\end{figure}

Figure \ref{fig:all_sample} shows the ratio of observed (not dust-corrected) $ L_{H\alpha}^{\prime}$ to $ L_{UV}^{\prime}$ as a function of stellar mass. (The prime sign on the $ L_{H\alpha}^{\prime}$, $ L_{UV}^{\prime}$, and $\xi_{ion}^{\prime}$ is to distinguish them as the not dust-corrected quantities.) Because the {\it Hubble} images are far more sensitive than our Keck/MOSFIRE observations, our primary incompleteness is determined by the depth of the spectroscopy. We are therefore concerned about completeness for galaxies with low $L_{H\alpha}^{\prime}$ and, thus, low $\xi_{ion}^{\prime}$. We therefore decide to only include galaxies in our final sample with spectra that are sensitive to the ``worst-case" (lowest) observed $L_{H\alpha}^{\prime}$ that can be expected. 

\begin{figure}
    \centering
    \includegraphics[width=1 \linewidth]{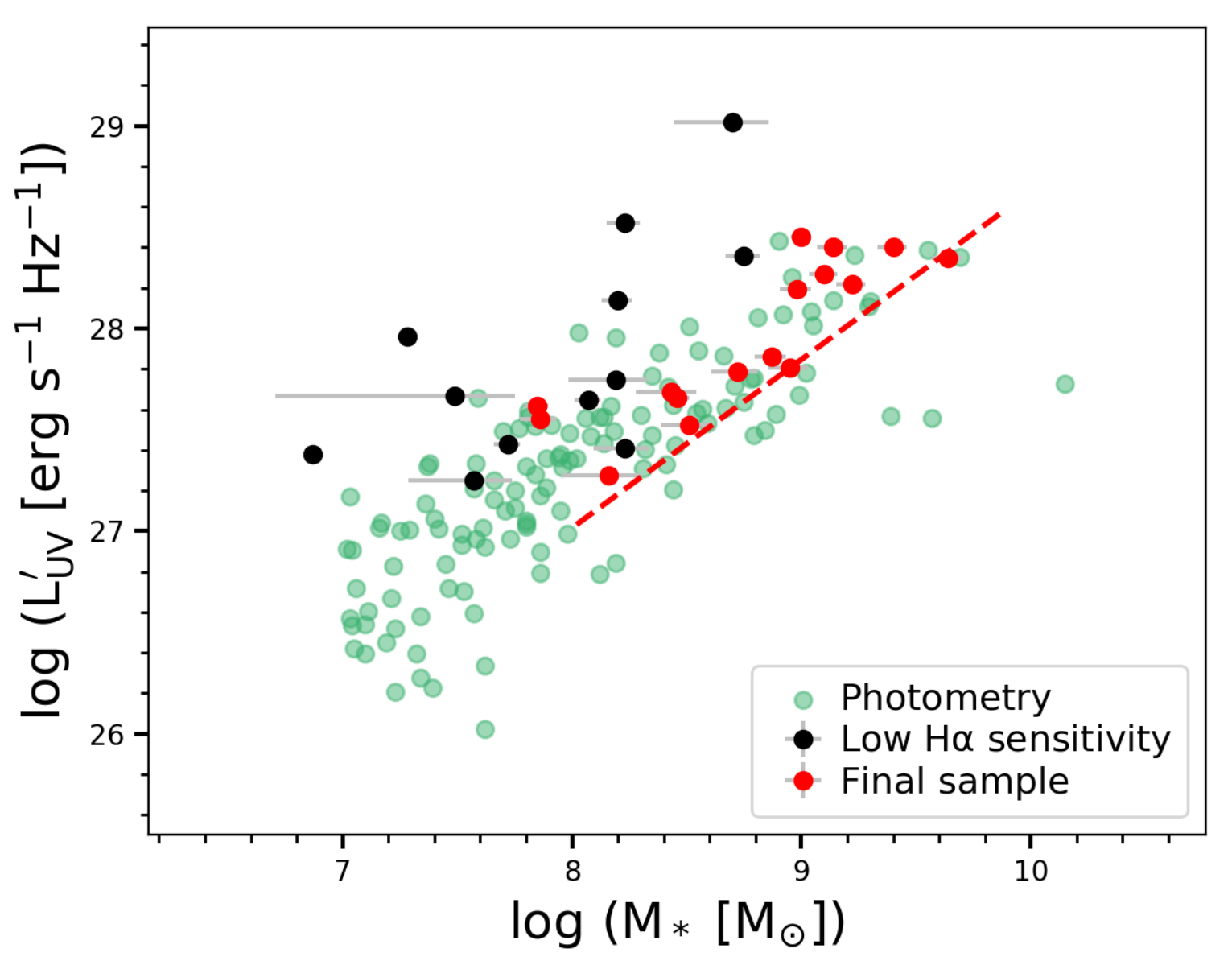}
    \caption{Not dust-corrected log ($L_{UV}^{\prime}$) - log (M$_*$) relation of our lensed galaxies. Green points are the parent photometric sample with $M_*$ above $10^7 M_{\odot}$. The final spectroscopic $\xi_{ion}^{\prime}$ sample is shown in red. Black points are removed from the $\xi_{ion}^{\prime}$ sample due to biases discussed in Section \ref{subsec:sample_select}. The galaxies in our final sample (red points) span a similar range in $L_{UV}^{\prime}$ at a given mass as the parent population, indicating that this sample is representative of low-mass galaxies at $1<z<3$. The red line denotes the lower edge of the log ($L_{UV}^{\prime}$) - log ($M_*$) main sequence trend of the parent sample that is used to exclude galaxies with insufficient sensitivity (black points) from the $\xi_{ion}^{\prime}$ sample described in Section \ref{subsec:non_dust_corr_xi}.}

    \label{fig:uv_mass}
\end{figure}

In order to determine the lowest $L_{H\alpha}^{\prime}$, we start by assuming the lowest $L_{UV}^{\prime}$ for the measured stellar mass of the galaxy. This is found with a line near the lower edge of the observed log($L_{UV}^{\prime}$)-log(M$_*$) relation (at $M_* > 10^{8}$ M$_{\odot}$ where our sample is complete) shown in Figure \ref{fig:uv_mass}. Once the worst-case $L_{UV}^{\prime}$ is determined, we then assume a worst-case log(\si) to find the faintest expected $L_{H\alpha}^{\prime}$. We determine this worst-case value at the higher masses ($>10^9 M_{\odot}$,  where we are complete), where we see that the lowest log($L_{H\alpha}^{\prime}/L_{UV}^{\prime}$) in our sample is $\sim13.2$. Finally, we compare this faintest $L_{H\alpha}^{\prime}$ to our line sensitivity (assumed as $3\sigma$ H$\alpha$ flux detection) to determine what magnification is required to detect H$\alpha$ in our spectra. We keep all galaxies in our sample that have a high enough magnification. In this way, we ensure that all galaxies remaining in our sample, have sufficiently sensitive spectra to detect galaxies with the lowest expected log(\si).

Once we find the magnification threshold at any given mass, we remove  galaxies in our sample whose magnifications are less than that threshold. There are 12 of these galaxies in our sample which are shown as black points in Figure \ref{fig:all_sample}. Now we only work with the remaining objects (red points) in our sample, which are not affected by any biases. We note that log($L_{H\alpha}^{\prime}/L_{UV}^{\prime}$) spans about one dex across the sample (13-14), as is evident in Figure \ref{fig:all_sample}.

We also perform a sanity check to determine whether our final sample can truly represent $\xi_{ion}^{\prime}$ in low-mass galaxies or suffers from any biases against low-mass faint galaxies. This investigation is primarily due to the fact that our spectroscopic sample is a magnitude-limited subsample of our parent photometric sample ($B<26.5$ AB). In this case, there is a possibility that we are populating the lower mass bins only with the most luminous and youngest galaxies and might be missing the faint sources. To ensure that our final sample does not suffer from this bias, we plot the log ($L_{UV}^{\prime}$) - log(M$_*$) distribution of our parent photometric sample and compare it to the final $\xi_{ion}^{\prime}$ sample in Figure \ref{fig:uv_mass}. This figure indicates that our final sample has a similar distribution to the parent sample, and is not biased toward high log($L_{UV}^{\prime}$) values at a fixed stellar mass down to the mass of $10^{7.8} M_{\odot}$. Hence, our final $\xi_{ion}^{\prime}$ sample is representative of low-mass galaxies at $1<z<3$ and is not biased against the low mass, faint galaxies.

\subsection{Dust Extinction Correction}
\label{subsec:dust}
We use the $A_V$ values derived from SED fits (Section \ref{subsec:sed}) and assume an SMC extinction curve to correct for the dust attenuation of the UV luminosity density. We also use the Balmer decrement ($L_{H\alpha}^{\prime}/L_{H\beta}^{\prime}$) to determine the $L_{H\alpha}^{\prime}$ attenuation assuming a Cardelli extinction curve \citep{Cardelli_1989}.

\section{Two Approaches to Flux Stacking for \si\ Estimates}
\label{sec:stacking}

Here we attempt to evaluate the representative log(\si) value of our sample. For this, we need to stack the dust-corrected $H\alpha$ and UV fluxes of individual galaxies. 
However, we note that the spread in log(\si) is large ($\sim$ 1 dex). Given such a large spread in the {\it logarithm} of \si, we need to be careful about how we stack, depending upon the question we are trying to answer. 

There are two \si\ values that we are interested in obtaining. First, we are interested in the properties of the {\it typical} galaxy, which can simply be obtained via the median, or the average of a symmetric distribution. Second, we are also interested in the total contribution of these galaxies to reionization, in which case we are interested in the {\it total} H$\alpha$ luminosity of all galaxies divided by the {\it total} UV luminosity of all galaxies. Such a number allows a direct conversion from UV luminosity functions to ionizing photon production rate densities. The stack in this case is {\it not} the average of the log(\si) values that many have calculated before. Instead, this stack is equivalent to an $L_{UV}$-weighted average of the $L_{H\alpha}/L_{UV}$ ratios of the galaxies, as shown below.

\begin{equation}
    \frac{\Sigma_i L_{H\alpha,i}}{\Sigma L_{UV,i}} = \frac{1}{\Sigma L_{UV,i}} \Sigma \frac{L_{UV,i}L_{H\alpha,i}}{L_{UV,i}}
\end{equation}

In order to obtain the composite log(\si) for each of these methods more quantitatively, we follow the procedures below. For the first method we take the average of the logarithms of the ratio of $L_{H\alpha}$ to $L_{UV}$ and refer to it as \one\ stacking method, and for the second method we take the ratio of the average $L_{H\alpha}$ to the average $L_{UV}$, then take the logarithm and refer to it as \two\ stacking method. 

These two methods will give different \si\ values for two reasons.  First, since the \one\ method takes the logarithm before averaging, it down-weights the importance of the high \si\ galaxies. Second, because the \two\ method is effectively an $L_{UV}$-weighted average of \si, it may differ from the average if there is a correlation between $L_{UV}$ and \si\ \citep[see for example ][]{Emami_2019}. The former method was used in \citet{Bouwens_2016a, Shivaei_2018} while the latter was used in \citet{ Matthee_2017, Lam_2019}. It is therefore important to account for these different stacking methods when comparing to previous works. 

We note here that our \two\ method does not strictly get the true value of the total, volume averaged log(\si) unless our sample galaxies also have similar luminosity and/or mass distributions as the true luminosity and/or mass functions. Of course, nearly all surveys of high-redshift galaxies have decreasing effective volumes at the faint-end of the survey, but this is especially true for lensing surveys, which also rely on rarer, high magnifications at the faint-end.  Though this will remain a concern, we show in Section \ref{sec:results} that \si\ does not change significantly with luminosity or mass, so this additional uncertainty is likely to be small.

In order to get the uncertainties in the composite log(\si) of each stacking method, we use the bootstrap resampling technique: for a data sample of size N, we draw N random values from the original sample and form a new sample of the same size and calculate its composite log(\si) the same way we did for the original sample. By repeating this 100,000 times, we build the distribution of the composite log(\si) values and calculate the 68\% confidence interval of this distribution as the uncertainty in the composite log(\si). We also incorporate the errors in the $H\alpha$, $H\beta$, and UV fluxes in this calculation by drawing a random value from a normal distribution with a width equal to the $1\sigma$ error for each flux. In this way, we include the $H\alpha$ and $H\beta$ flux errors on the $A_{H\alpha}$ determination and the $H\alpha$ and UV flux errors on the log (\si) determination.

\begin{deluxetable*}{cccc}
\tablecaption{log (\si) derived for different sub-samples and different stacking methods}
\tablenum{1}

\tablehead{\colhead{Subsample$^a$} & \colhead{\it{\one} $log(\xi_{ion})^b$} & \colhead{\it{\two} $log(\xi_{ion})^c$} & \colhead{No. of objects} } 

\startdata
$7.8<log(M_*)<8.8$ & $25.17\substack{+0.13\\-0.19}$ & $25.34\substack{+0.12\\-0.15}$ & 6 \\
$8.8<log(M_*)<9.8$ & $25.13\substack{+0.21\\-0.19}$ & $25.39\substack{+0.14\\-0.18}$ & 9 \\
$-22<M_{UV}<-19.5$ & $25.27\substack{+0.13\\-0.16}$ & $25.47\substack{+0.10\\-0.11}$ & 9 \\
$-19.5<M_{UV}<-17.3$ & $25.16\substack{+0.14\\-0.18}$ & $25.47\substack{+0.12\\-0.15}$ & 11 \\
$-2.4<\beta<-1.75$ & $25.15\substack{+0.14\\-0.23}$ & $25.46\substack{+0.11\\-0.15}$  & 12\\
$-1.75<\beta<-0.93$ & $25.27\substack{+0.13\\-0.17}$ & $25.45\substack{+0.13\\-0.16}$ & 8 \\
\enddata

\tablenotetext{a}{Log (\si) measured for different sub-samples of log($M_*$) (in unit of $M_{\odot}$), UV magnitude, and UV continuum slope.}
\tablenotetext{b}{log(\si) inferred from ``\one'' stacking method.}
\tablenotetext{c}{log(\si) inferred from ``\two'' stacking method.}
\label{table:1}

\end{deluxetable*}

\section{Results}
\label{sec:results}

In this section we discuss the relationship between \si\ and stellar mass as well as other physical quantities such as UV magnitude, UV spectral slope ($\beta$), and the equivalent widths of nebular emission lines and compare that with other studies. In Table \ref{table:1}, we present the composite log(\si) and its error in bins of stellar mass, UV magnitude ($M_{UV}$), and UV continuum slope ($\beta$) obtained from the two stacking methods described in Section \ref{sec:stacking}. 

\subsection{\si\ vs. Stellar Mass}
\label{subsec:xi_mass}

\begin{figure*}
    \centering
    \includegraphics[width=1 \linewidth]{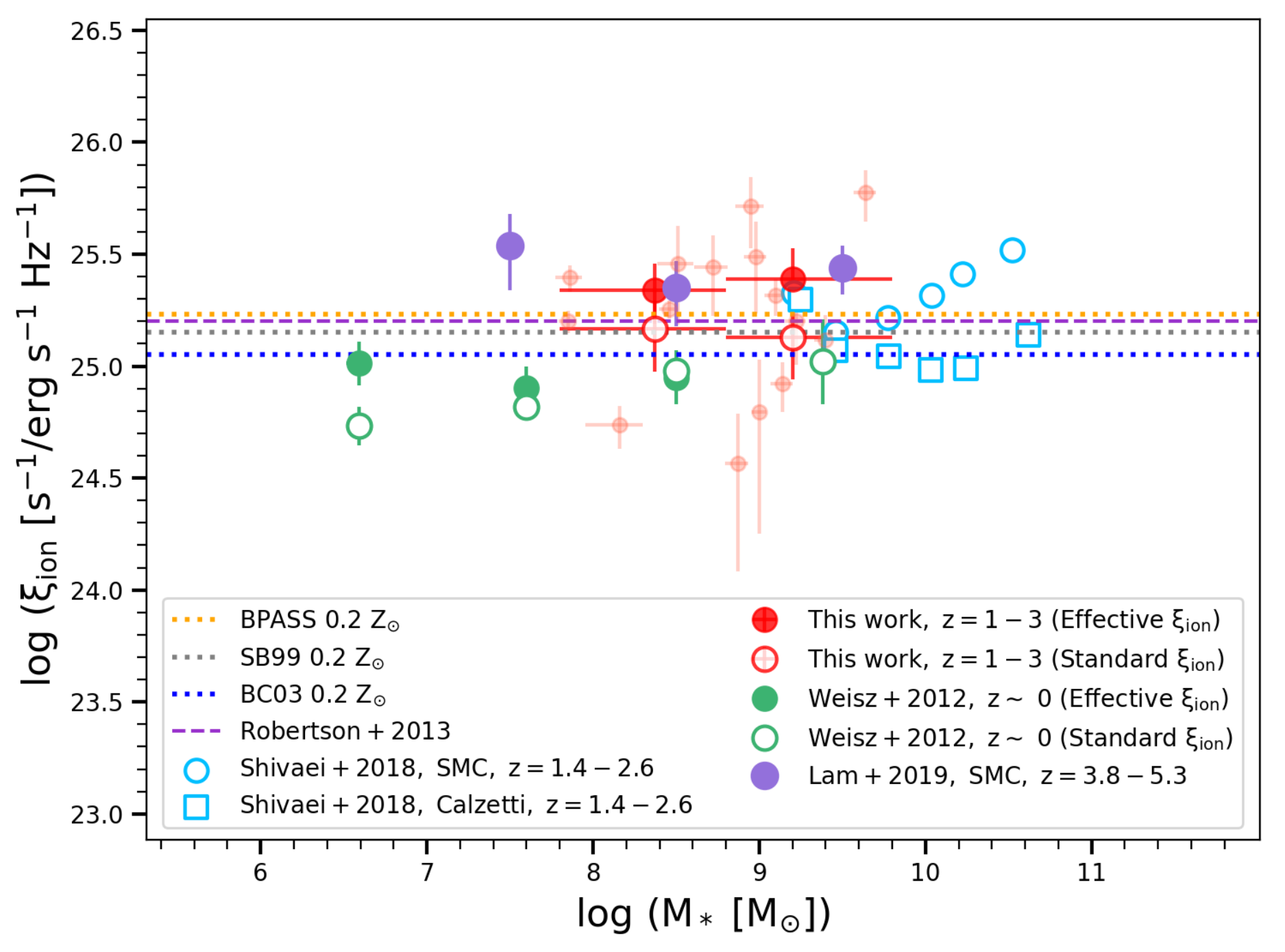}
    \caption{Log(\si) as a function of log($M_{*}$). Log(\si) derived from the \one\ stacking method are shown in red open circles and the \two\ stacking method in red filled circles. The log(\si) inferred from \two\ method is $\sim$ 0.2 dex larger than that of the \one\ method.
    Green open and filled circles denote the local sample of \citet{Weisz_2012} applying the \one\ and \two\ stacking methods respectively. Sky blue open squares and circles denote the MOSDEF sample \citep{Shivaei_2018} of higher stellar mass galaxies using \citet{Calzetti_2000} and SMC \citet{Gordon_2003} UV dust corrections respectively. Purple circles show the \citet{Lam_2019} sample of faint ($L_{UV}< $  0.2 $L_*$) galaxies at higher redshifts ($z=3.8-5.3$). For a better comparison of samples with similar stacking methods, we use open markers to indicate the \one\ stacking method and filled markers to indicate the \two\ stacking method. The dashed line is the canonical value of 25.2 from \citet{Robertson_2013}. The local sample of \citet{Weisz_2012} indicates lower log(\si) compared to ours. High-redshift samples of \citet{Shivaei_2018} and \citet{Lam_2019} lie within the 1$\sigma$ error bars of our two stacking methods.
    Orange, gray, and blue lines indicate the log (\si) predicted by three different single stellar models, (BPASS model \citep{Eldridge_2017}, Starburst99 \citep{Leitherer_2014}, and BC03 \citep{Bruzual_2003}), assuming a constant star formation history and $0.2 Z_{\odot}$ metallicity.}
    \label{fig:xi_mass_Weisz_Shivaei}
\end{figure*}

Galaxy stellar mass ($M_*$) is known to correlate with metallicity, which affects the stellar temperatures and, thus, \si. We are therefore interested in examining the dependence of log(\si) on stellar mass for our sample.
We present the log(\si) derived from our two stacking methods as a function of log(stellar mass) in Figure  \ref{fig:xi_mass_Weisz_Shivaei}. 
 
As can be seen in Figure \ref{fig:xi_mass_Weisz_Shivaei}, the composite log(\si) is higher by at least $0.2$ dex at all mass bins when using the \two\ method compared to the \one\ method. Since the errors in log(\si) of some galaxies are not negligible compared to the size of the spread in the log(\si) distribution of the sample, it might be thought as if the stacked log(\si) derived from \two\ method is perhaps higher because of the large noise in these galaxies.
We also check to make sure this enhancement is primarily due to the intrinsically high luminosities and not the noise. For that, we need to know how much the noise from our measurement has spread our observed log(\si) distribution. We run a simple simulation here: We first construct a normally distributed log(\si) of 1000 sources with an intrinsic spread of $\sigma_{int}$ and perturb them with the fractional noise which is randomly drawn from the errors in \si\ of our sample. We then calculate the spread in this simulated \si\ distribution as $\sigma_{sim}$. In order for the simulated spread to be equal to the observed spread (0.35), the intrinsic spread is required to be $\sim0.29$ dex, which implies $\sim0.19$ dex spread due to noise. Thus, the intrinsic spread is larger and is the primary reason for the increased log(\si) enhancement calculated via the \two\ stacking method.

Now we compare our results with other studies at different redshifts or different stellar masses. First we compare to a sample of local low-mass galaxies from \citet{Weisz_2012}. We have determined the composite log (\si) of this sample in four mass bins, using the same two stacking methods we used for our sample, shown as green points in Figure \ref{fig:xi_mass_Weisz_Shivaei}. Similar to our sample, we see that the log(\si) measured from the \two\ method is similar to or higher than the one derived from the \one\ method in this sample. In particular, the difference between the two methods increases at lower masses where the scatter in the log (\si) is dramatic and is likely due to the increasing burstiness, as was found by \citet{Emami_2019}.

Comparing our results with \citet{Weisz_2012}, we find that at a given mass, our sample shows higher log(\si) relative to that of \citet{Weisz_2012}  (compare red markers with green ones), suggestive of a log(\si) evolution with redshift in the low-mass systems. We discuss possible explanations for this in Section \ref{sec:discussion}.

We also compare our sample with higher mass galaxies at similar redshift ($1.4<z<2.6$) from the MOSDEF Survey \citep{Shivaei_2018}. In Figure \ref{fig:xi_mass_Weisz_Shivaei} we show the log(\si) values for MOSDEF assuming SMC \citep{Gordon_2003} and \citet{Calzetti_2000} UV dust extinction corrections.

We see that the log(\si) of our sample is in good agreement with that of \citet{Shivaei_2018} at $10^{9}- 10^{9.5} M_{\odot}$ within $1\sigma$ uncertainty, in the mass range where the two samples overlap. In fact, the log(\si) values of our galaxies in our sample and those at higher stellar mass are consistent at all stellar masses. Thus, there is no evidence for a trend in log(\si) with stellar mass from $10^{7.8}- 10^{11} M_{\odot}$.

We also compare to the high redshift sample of \citet{Lam_2019}, shown as purple circles in Figure \ref{fig:xi_mass_Weisz_Shivaei}. The sample is at redshift $3.8<z<5.3$. Galaxies in this sample are primarily selected to have Ly$\alpha$ emission in the MUSE data. The sample includes galaxies of faint UV luminosities $-20.5<M_{UV}<-17.5$, similar to the galaxies in our intermediate-redshift sample. The log(\si) is inferred from the $H\alpha$ equivalent width which in turn is derived from a power-law model spectrum fit through the flux of stacked {\it Spitzer}/IRAC [3.6]-[4.5] bands. The derived $H\alpha$ is then divided by the stacked UV fluxes. To that end, their way of log(\si) determination is similar to our \two\ stacking method. The log(\si) obtained from the \two\ method in our sample is consistent with that of \citet{Lam_2019} within 1$\sigma$ error (compare red and purple filled markers).

\subsection{\si vs. UV Absolute Magnitude}
\label{subsec:xi_Muv}

$M_{UV}$ is one of the easiest observables to obtain for high redshift galaxies. Furthermore, the integral of the UV luminosity function is a critical calculation in determining the ionizing emissivities of galaxies. Therefore, we are particularly interested in whether or not there is any correlation between $M_{UV}$ and \si.

\begin{figure*}
    \centering
    \includegraphics[width=1 \linewidth]{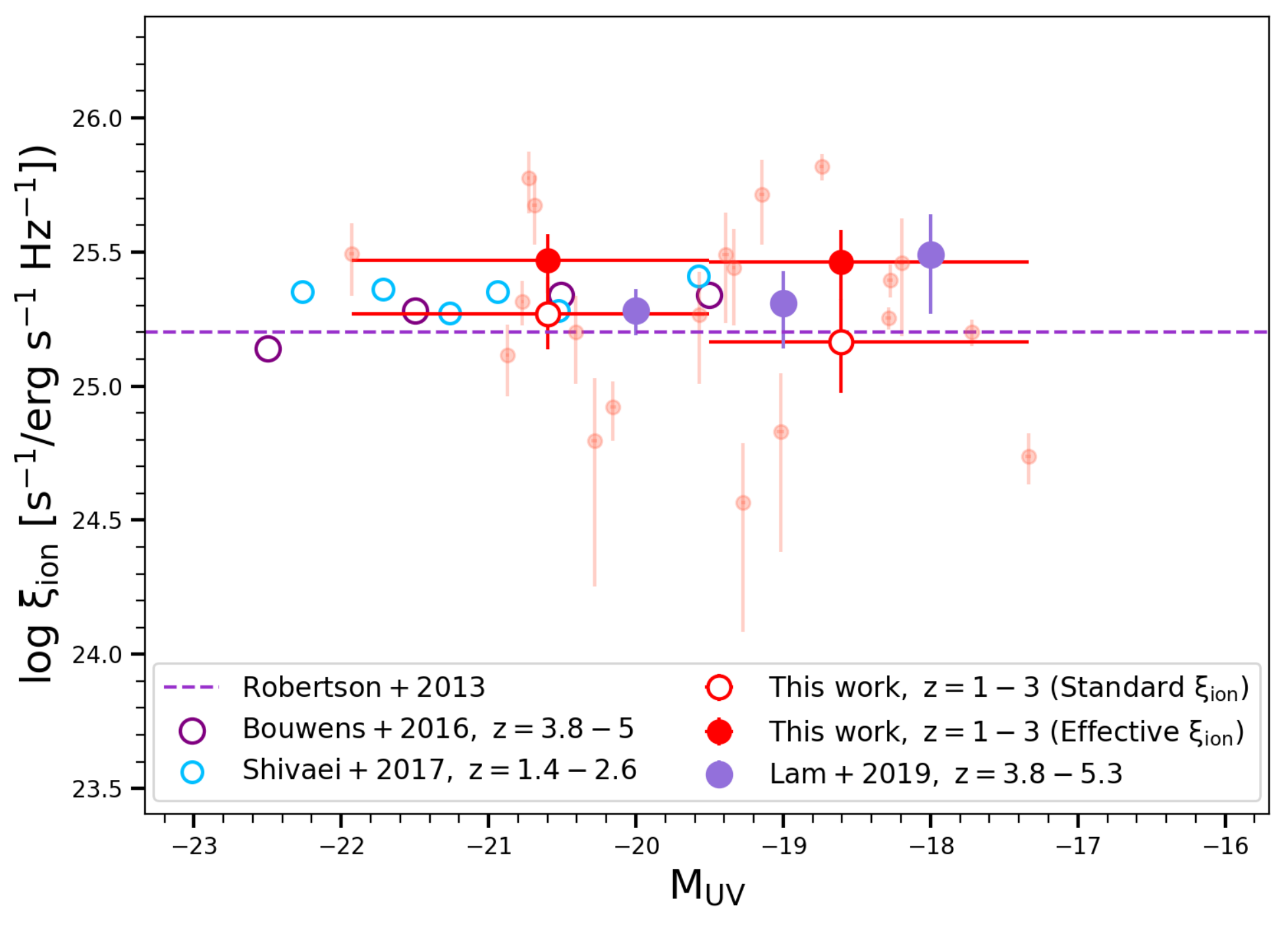}
    \caption{Log(\si) as a function of UV magnitude, $M_{UV}$. Small, light red circles denote individual galaxies in our sample. The large open and filled red circles show the log($\xi_{ion}$) derived from the \one\ and \two\ stacking methods, respectively. Sky blue circles show the stacks from \citet{Shivaei_2018} for more massive $z \sim 2$ galaxies and an SMC UV dust correction. Dark and light purple circles denote \citet{Bouwens_2016a} and \citet{Lam_2019} samples at $z \sim 4-5$ respectively. Similar to Figure \ref{fig:xi_mass_Weisz_Shivaei}, for a better comparison of samples with similar stacking methods, we use open markers to indicate the \one\ stacking method and filled markers to show the \two\ method. Our values agree with other studies within 1$\sigma$ significance when the same stacking method as ours are used. No significant dependence of log(\si) with $M_{UV}$ is found.}
    \label{fig:xi_Muv}
\end{figure*}

In Figure \ref{fig:xi_Muv} we plot log(\si) as a function of $M_{UV}$.
We determine log(\si) for two bins of $M_{UV}$ ($-22 <M_{UV}<-19.5$ and $-19.5 \leq M_{UV}< -17.2$) using the \one\ and \two\ stacking methods as were described in section \ref{sec:stacking}.

As in the previous section, we also need to take care that we only include galaxies for which we could detect very low $L_{H\alpha}$. However, in this case, we are sampling galaxies based on their $M_{UV}$, so we don't need to add a step of assuming an $M_{UV}$-$M_*$ relation. Instead, we simply determine which galaxies could be detected if they had the very low observed log($L_{H\alpha}/L_{UV}) \gtrsim 13.2$ and use each galaxy's measured $L_{UV}$. 

As was mentioned earlier in Section \ref{subsec:sample_select}, we return galaxies with high stellar mass errors to our sample as their masses are irrelevant in this analysis.

Log(\si) derived from the \one\ method is similar in the two $M_{UV}$ bins (25.17 and 25.28); while the \two\ method gives a log(\si) of 25.47 for both bins.  In both $M_{UV}$ bins, the \two\ method gives log(\si) values $\sim 0.2$ dex larger than that of the \one\ method. We also show results from \citet{Shivaei_2018, Bouwens_2016a}, and \citet{Lam_2019} in Figure \ref{fig:xi_Muv}.
Comparing log(\si) of all works with analogous stacking techniques, we find that our values are in agreement with other works within 1$\sigma$ significance.

We do not find any evidence of significant dependence of log(\si) on $M_{UV}$ in our sample, in agreement with these other studies.

\begin{figure*}
    \centering
    \includegraphics[width=1 \linewidth]{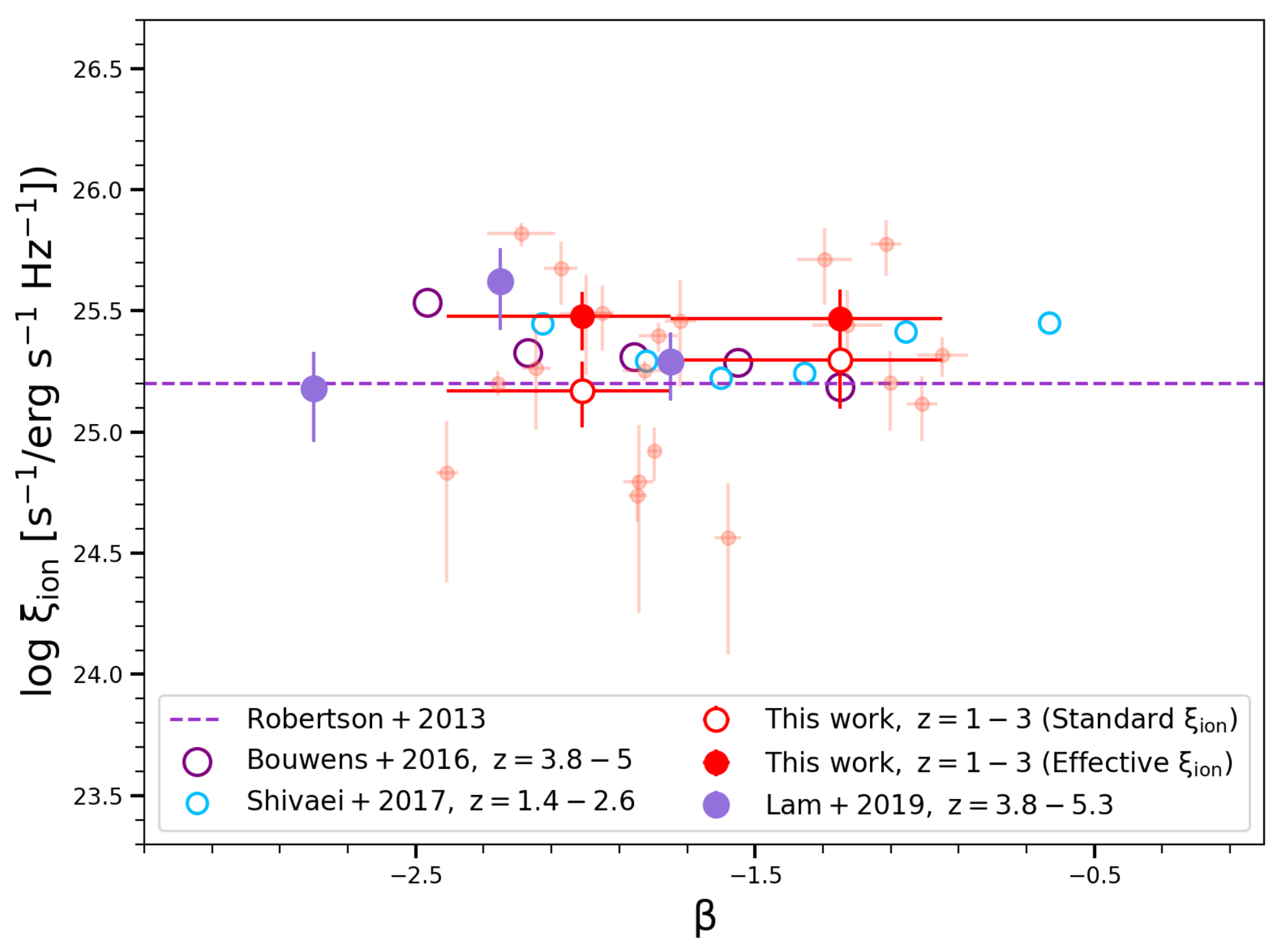}
    \caption{Log(\si) as a function of UV slope $\beta$. The symbols are the same as in Figure \ref{fig:xi_Muv}. No dependence of log(\si) with $\beta$ is seen in our sample.}
    \label{fig:xi_beta}
\end{figure*}

\subsection{\si\ vs. UV Continuum Slope}
\label{subsec:xi_beta}

The UV continuum slope, $\beta$, is related to both the metallicity and age of the stellar populations and therefore, the inferred ionization capability of a galaxy driven by its young star populations. Therefore, we investigate if \si\ is correlated with the more easily observable $\beta$. 

 The individual log(\si) values vs. $\beta$ are plotted in Figure \ref{fig:xi_beta}. We split the sample into two bins of $\beta$ ($-2.4<\beta<-1.75$ and $-1.75 \leq \beta<-0.9$) and apply the same two stacking methods at each bin as we used for log($M_*$) and $M_{UV}$. We find a similar log(\si) of 25.45 and 25.47 for the \two\ method and a log(\si) range of 25.18-25.3 for the \one\ method. We do not see any evidence for log(\si) being correlated with $\beta$ in our sample. Again we find that our log(\si) stack values are consistent with those of other studies at similar or higher redshifts.

\subsection{\si\ vs. {$\mathrm{EW_{[OIII]\lambda 5007}}$, $\mathrm{EW_{H\alpha}}$}}
\label{subsec:xi_EW}

Finally, we investigate the relationship between log($\xi_{ion}$) and the equivalent widths of optical nebular emission lines. One expects a positive correlation because the optical line equivalent widths are directly related to the luminosity-weighted age of the stellar populations, which itself affects \si. 

\begin{figure}
    \centering
    \includegraphics[width=1 \linewidth]{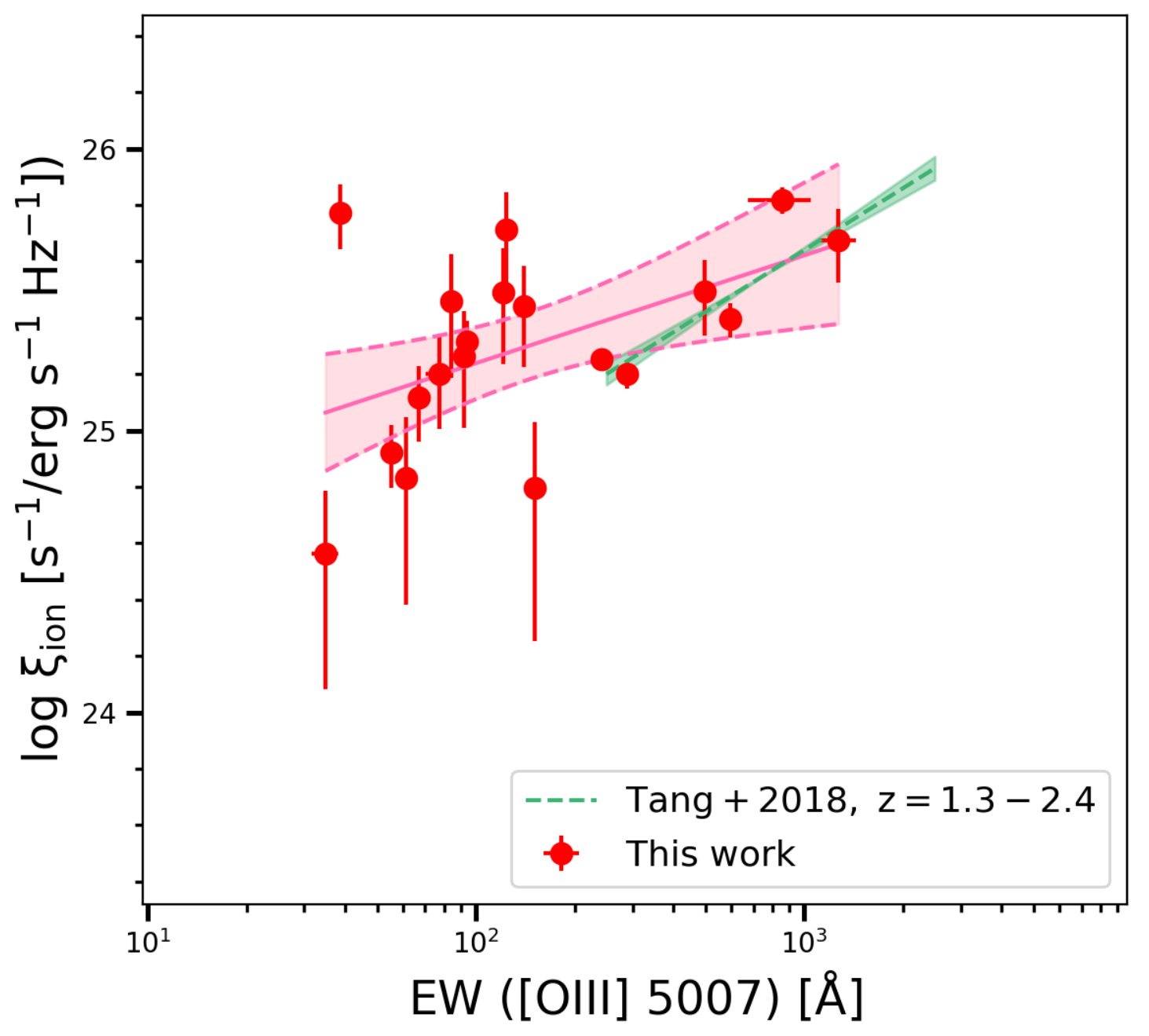}
    \includegraphics[width=1 \linewidth]{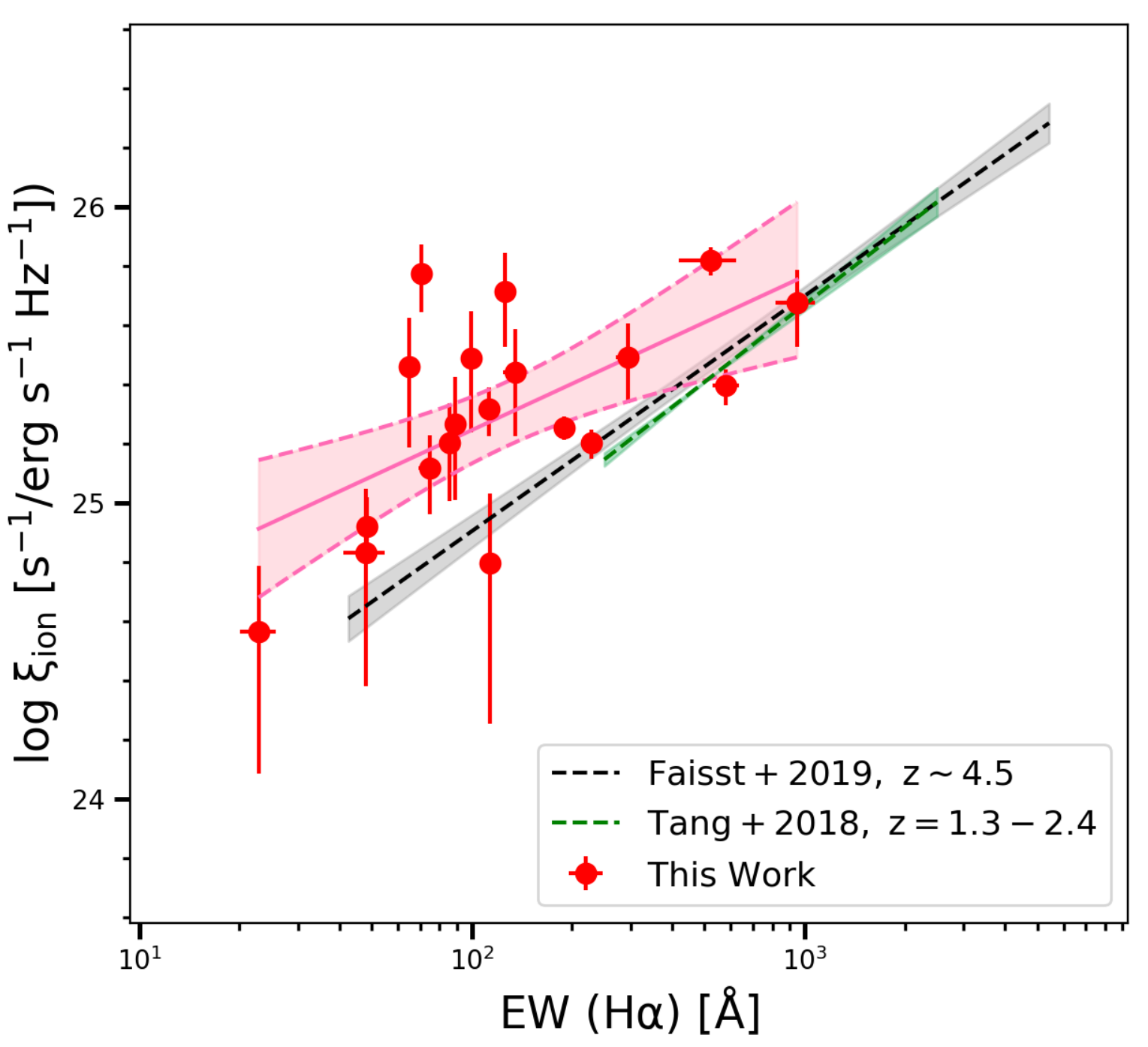}
    \caption{Top:log (\si) vs. \oiii\ 5007 equivalent width. The solid red line and the pink region denote the best-fit line and $1\sigma$ confidence region respectively. The green dashed line is from \citet{Tang_2019} for extreme $[OIII]$ emitters at $1.3<z<2.4$. Overall, a positive slope of $0.38 \pm 0.16$ is apparent between the two properties, but less steep than \citet{Tang_2019}. Bottom: log (\si) vs. H$\alpha$ equivalent width ($EW_{H\alpha}$). There is a slope of $0.52 \pm 0.16$ between the two properites. The gray line denotes the \citet{Faisst_2019} relation at $z\sim 4.5$, which overlaps with the \citet{Tang_2019} (green) but extends to a larger range of $H\alpha$ equivalent widths (40-5000 $\AA$).}
    \label{fig:EW}
\end{figure}

This relationship was first investigated by \citet{Chevallard_2018}, who found that log(\si) in galaxies with strong ionizing emissivities are scaled with the equivalent width of the combined \oiii\ 4959,5007 lines. They showed this for a sample of local star-forming galaxies with very high rest-frame equivalent widths ($560<EW_{[OIII]\lambda 5007}<2370\ \AA$). \citet{Tang_2019} confirmed the existence of such a scaling relation for a sample of 227 low-mass ($10^7<M_*/M_{\odot}<10^{10}$), \oiii\ emitters with $225<EW_{[OIII]\lambda 5007}<2500\ \AA$ at $1.3<z<2.4$, suggesting that higher equivalent width systems are more efficient ionizing agents. Given that, we aim to test this for our galaxies to see if this relation further extends to lower equivalent width systems or not.
We calculate the $EW_{[OIII]\lambda 5007}$ by taking the ratio of the [OIII] emission line flux to the flux of the rest-frame 5007 $\AA$ continuum from our {\it HST} near-IR fluxes, which have been corrected for emission line contamination.
We show log(\si) vs. log \oiii5007 equivalent width (EW$_{[OIII]\lambda 5007}$) in the top panel of Figure \ref{fig:EW}.
Our galaxies span a large range of rest-frame equivalent widths ($20<EW_{[OIII]\lambda 5007}<1500\ \AA$), but generally extend lower than these previous studies. There is a trend of increasing log(\si) with log($EW_{[OIII]\lambda 5007}$).
To quantify this trend, we fit a line to the sample using ordinary least squares and plot the best fit, along with the $68\%$ confidence region. We see a correlation between log($\xi_{ion}$) and log(EW$_{[OIII]\lambda 5007}$) with a slope of $0.38 \pm 0.16$. 
 
In addition we overlay the trend from \citet{Tang_2019} at larger $EW_{[OIII]\lambda 5007}$ which is steeper than ours, with smaller uncertainty in the fit. We note that our galaxies at $EW_{[OIII]\lambda 5007}> 200\ \AA$ display a similar trend to that of \citet{Tang_2019}.
The discrepancy between the two trends suggests that the slope in the log(\si)-log($EW_{[OIII]\lambda 5007}$) gets shallower at lower equivalent widths.

We also plot the log(\si) vs. H$\alpha$ equivalent width ($EW_{H\alpha}$) relation in the bottom panel of Figure \ref{fig:EW}. After fitting a line through the points, we find a significant correlation, with a slope of $0.52 \pm 0.16$ between the two indicators. We further overplot the trend from \citet{Tang_2019} and \citet{Faisst_2019} which contains galaxies at $z\sim 4.5$ and stellar masses $>10^{9.7} M_{\odot}$. Our trend has a similar slope to those of \citet{Tang_2019} and \citet{Faisst_2019} but again with a larger uncertainty in the fit. Such a steep slope implies that log(\si) is more correlated with $EW_{H\alpha}$ than with $EW_{[OIII]\lambda 5007}$, as was reported by \citet{Tang_2019}.
In addition, \citet{Reddy_2018} have also found similar trends of $EW_{H\alpha}$ and $EW_{[OIII]\lambda 5007}$ vs. \si\ to ours for more massive galaxies in the MOSDEF survey ($10^9<M_{*}/M_{\odot}<10^{10.5}$) at $1.4<z<3.8$.
\citet{Tang_2019} argue that the log(\si)-$EW_{[OIII]}$ and log(\si)-$EW_{H\alpha}$ correlation should not hold at lower equivalent widths (below $200\ \AA$). According to \citet{Tang_2019}, the EWs correlate with \si\ only within the first 100 Myrs since the onset of star formation. After this time, both young (O-type) and intermediate-aged (B- and A-type) populations reach equilibrium, resulting in a constant $L_{H\alpha}$-to-$L_{UV}$ ratio and a plateau in \si\ versus EW. This is also evident in our sample as we get shallower slopes when including lower EWs into the line fits (below 200$\AA$). However, this star formation history interpretion is only correct if one assumes a constant star formation history.
A more comprehensive investigation of the \si\ dependence on the EWs requires additional analysis of the star formation histories of galaxies as well as other physical properties, which is beyond the scope of this paper.

Finally, we note that this correlation between log(\si) and the equivalent widths of some ionization-sensitive nebular emission lines can be used as a proxy for \si\ at high-redshifts when the direct measurement of rest-frame \uv\ is not available \citep{Chevallard_2018, Tang_2019}.

\section{Discussion}
\label{sec:discussion}

In Section \ref{subsec:xi_mass} we reported an increase in the log(\si) of our $1.4<z<2.7$ sample relative to the low-redshift sample of \citet{Weisz_2012} (See Figure \ref{fig:xi_mass_Weisz_Shivaei}).
This suggests that at higher redshifts, galaxies with mass range of  $10^{7.8} \leq M_{\odot} \leq 10^{9.5}$ produce more ionizing photons relative to the non-ionizing UV photons when compared to their low redshift counterparts.
Here we provide possible explanations for this difference between high- and low-redshift samples.

First, the oxygen-to-iron abundance ratio of galaxies affects the production of ionizing photons at high redshift.
Recent studies by \citet{Steidel_2016, Strom_2017} show that in high-mass ($9 \leq log(M_{*}/M_{\odot})\leq 10.8$), high-redshift galaxies ($z=2.4 \pm 0.11$), the [O/Fe] abundance is super-solar ($\simeq 4-5$ [O/Fe]$_{\odot}$) referred to as ``$\alpha$ enhancement." This has been shown in the composite UV spectrum of a representative sample of galaxies in KBSS-MOSFIRE spectroscopic survey \citep{Steidel_2014}. They found that emission spectra from photoinization modeling best matches their composite UV spectra with stellar models with low stellar metallicities ($Z/Z_{\odot} \sim 0.1$), while the gas-phase oxygen abundances measured from nebular emission lines are $\sim4$ times higher. Given that stellar opacity is dominated by iron, this suggests a super-solar [O/Fe].

The deficit of iron in high-redshift galaxies can be explained by a model in which iron is produced during a delayed detonation of type Ia supernovae (SNe) \citep{Khokhlov_1991}.
As a consequence, in high-redshift galaxies not all white dwarf stars have detonated and released iron into the interstellar medium (ISM). Since iron predominantly controls the opacity of stellar atmospheres, its deficiency allows stars of a given mass to be hotter and, thus, have higher ionizing photon production at higher redshifts, leading to an increase in \si\ compared to local samples. It is likely the case that stellar populations of lower mass galaxies at $z\sim2$ are as young as the higher mass galaxies of \citet{Steidel_2014}, and therefore exhibit a similar $\alpha$ enhancement. To confirm this requires measurement of the iron abundance of low-mass galaxies via absorption lines in their UV spectrum. 

Second, this excess in the ionizing UV photons could be due to a recent increase in the star formation activity of high-redshift galaxies resulting in an enhancement in the L$_{H\alpha}$ relative to the \uv.
This effect has also been reported in \citet{Faisst_2019} such that their $z\sim 4.5$ main-sequence galaxies indicate a \si\ median of 25.5 which is 0.3 dex above the typically used canonical value of \citet{Robertson_2013}.
This recent star formation activity can be in the form of continuous increase in the star formation history of the galaxy, which is typical in $z>4$ galaxies \citep{Behroozi_2019}, or a recent, rapid burst. Either of these star formation scenarios will lead to an increase in the number density of young stellar populations relative to the number density of intermediate-aged stellar populations in galaxies and ultimately results in an excess in the L$_{H\alpha}$ to \uv\ ratios. However, exploring the effect of star formation variation on the ionizing photon production efficiency requires a deeper analysis of the star formation properties of galaxies at different epochs, which is beyond the scope of this paper and will be the subject of a future investigation.

We also investigate the \si\ predicted by different star formation synthesis models, BPASS \citep{Eldridge_2017}, BC03 \citep{Bruzual_2003}, and Starburst99 \citep{Leitherer_1999, Leitherer_2014} for a constant star formation history (which should be equivalent to the average of many galaxies at various stages of burstiness) and compare them with our observed values as shown in Figure \ref{fig:xi_mass_Weisz_Shivaei}. Assuming a 0.2 $Z_{\odot}$ metallicity, Chabrier IMF \citep{Chabrier_2003}, and Padova isochrone \citep{Bertelli_1994, Bressan_1993, Fagotto_1994}, we find that these models produce log(\si) values within 25-25.2. When including the effect of stripped, binary stars to the BPASS \citep{Eldridge_2017} and Starburst99 \citep{Gotberg_2019} single star models, we find only a small, $\sim5$\%, enhancement to \si. This is because these stripped, binary stars emit HI-ionizing photons at a rate which is $5\%$ of the rate of HI-ionizing photons emitted by the massive single O-type stars \citep{Gotberg_2019}. As a result,  when a constant star formation history is assumed, the emission from the massive single stars always dominates the emission from other less massive stripped, binary stars. Therefore, the evolution in \si\ can not fully be explained by an evolution in binarity as predicted by these models.

\section{Summary}
\label{sec:summary}

In this paper we measure the ionizing photon production efficiency per unit 1500 $\AA$ UV luminosity, \si, of a sample of low-mass ($7.8<log(M*/M_{\odot})<9.8$) lensed galaxies at $1.4<z<2.7$. We obtained rest-frame optical spectroscopy of these faint sources that are magnified by the foreground lensing clusters Abell 1689, MACS J0717, and MACS J1149, enabling us to extend the \si\ measurement to lower masses and fainter UV magnitudes ($M_{UV}<-18$) than previously probed at these redshifts. We use the ratio of the $H\alpha$ luminosity (from Keck/MOSFIRE spectroscopy) and the 1500 $\AA$ UV luminosity density (from {\it HST} imaging) to measure \si. We limit our sample to those objects where we are complete in our measurement of \si.

We divide the sample into bins of different physical quantities such as stellar mass, absolute UV magnitude ($M_{UV}$) and UV spectral slope ($\beta$) and calculate the stacked log(\si) in each bin using two different stacking methods.
The most common method is to take the average of the log ($L_{H\alpha}/L_{UV}$) of galaxies to determine the standard log(\si) value, referred to as the ``\one" stacking method. The second method is to take the log of sum($L_{H\alpha}$)/sum($L_{UV}$) which we refer to as the ``\two" stacking method. This method is preferable when one is interested in calculating the total ionizing UV luminosity density from the non-ionizing UV luminosity function. Here we list our main results:
    
\begin{itemize}

\item In samples with a large spread in the log(\si) distribution, the stacked log(\si) from the two stacking methods can be significantly different. This is evident in the low mass local sample of \citet{Weisz_2012} in Figure \ref{fig:xi_mass_Weisz_Shivaei}.

\item  We measure a value of log(\si) $\sim25.47\pm 0.09$ for our UV-complete sample in the range $-22<M_{UV}<-17.3$ and $\sim25.37\pm0.11$ for our mass-complete sample in the range $7.8<\text{log}(M_*)<9.8$. The slight difference between these two values is due to small differences in the samples.

\item We find that the log(\si) derived from \two\ method is about 0.2 dex higher than that of the \one\ method in our sample of $z\sim 2$ galaxies, meaning that low UV luminosity systems may contribute $\sim60$\% more ionizing photons than inferred from other stacking methods. 

\item The measured log(\si) of our $z\sim2$ sample is higher than the low mass local sample of \citet{Weisz_2012} by $\sim0.2-0.3$ dex when measured in a consistent manner. We argue that this can be attributed to different physical properties in high- and low-redshift galaxies:
i) Delayed Type Ia supernovae results in an $\alpha$-enhancement (lower Fe relative to O) in the stellar population, which causes stars of a given mass to be hotter and, thus, have higher ionizing photon production \citep{Steidel_2016}. 
ii) An increase in the recent star formation activity of the high-redshift galaxies can also increase the relative number of young stars, thereby increasing the ratio of ionizing photons to non-ionizing photons.

\item We find similar \si\ values to galaxies of higher mass at similar redshift \citep{Shivaei_2018} and similar mass at higher redshift \citep{Lam_2019}. \si\ derived from these three samples are roughly consistent with the predictions of the BPASS binary stellar models with an assumption of 0.2 $Z_{\odot}$ stellar metallicity.

\item  We find no strong dependence between \si\ and $M_{UV}$ or UV spectral slope, $\beta$, consistent with \citet{Bouwens_2016a, Shivaei_2018, Lam_2019}.

\item There is a positive correlation between \si\ and both $H\alpha$ and [OIII]$_{5007}$ equivalent widths in our faint, lower equivalent width systems. This confirms that the equivalent width of these strong optical lines can act as a proxy for \si, though the relation appears to be less steep and with larger scatter at lower equivalent widths. 

\item  We find an intrinsic scatter of $\sim 0.29$ dex in the log(\si) distribution of our sample. Many physical factors can cause this scatter. In a future paper we will investigate the underlying causes of this scatter in our lensed, high-redshift sample. 

\end{itemize}

We thank the anonymous referee for providing useful comments
that helped improve the quality of this paper.
This material is based upon work supported by the
National Science Foundation under Grant No. 1617013.
Support for programs No. 12201,12931, 13389 and 14209 was provided by NASA through a grant from the Space Telescope Science Institute, which is operated by the Association of Universities for Research in Astronomy, Inc., under NASA contract NAS5-26555.

The authors wish to recognize and acknowledge the very significant cultural role and reverence that the summit of Maunakea has always had within the indigenous Hawaiian community. We are most fortunate to have the opportunity to conduct observations from this mountain.

Facilities: Keck:I (MOSFIRE), HST (WFC3,
ACS).
D.R.W. acknowledges fellowship support from the Alfred P. Sloan Foundation and the Alexander von Humboldt Foundation.

\bibliography{xi_ion_in_dwarf_galaxies_revised_II}

\end{document}